# High quality factor single-crystal diamond mechanical resonators

P. Ovartchaiyapong, L. M. A. Pascal, B. A. Myers, P. Lauria, A. C. Bleszynski Jayich

Department of Physics, University of California Santa Barbara, Santa Barbara CA, 93106 USA

Single-crystal diamond is a promising material for MEMs devices because of its low mechanical loss, compatibility with extreme environments, and built-in interface to high-quality spin centers. But its use has largely been limited by challenges in processing and growth. We demonstrate a wafer bonding-based technique to form diamond on insulator, from which we make single-crystal diamond micromechanical resonators with mechanical quality factors as high as 338,000 at room temperature. Variable temperature measurements down to 10 K reveal a nonmonotonic dependence of quality factor on temperature. These resonators enable integration of single-crystal diamond into MEMs technology for classical and quantum applications.

Traditionally renowned for its excellent mechanical, thermal, and optical properties[1-3] single-crystal diamond (SCD) has also recently demonstrated promise in electronic[4] and quantum applications[5-10]. Mechanical systems, such as microelectromechanical (MEMs) and nanoelectromechanical (NEMs) devices, could exploit diamond's exceptional characteristics in a wide variety of applications, ranging from ultrasensitive force detection[11,12] to quantum optomechanics[13-16]. For instance, proposals to mechanically couple NEMs devices to intrinsic quantum two-level systems [17,18] could be realized in a diamond NEMs containing nitrogen vacancy (NV) centers in diamond. NV centers are diamond defects with electronic spins displaying long (> 1 ms in bulk) quantum coherence times at room temperature, and they are being actively pursued for applications in quantum information and sensing[6-8,19]. A nanomechanical-spin interface could enable mechanical control of spin states as well as provide a hybrid approach to a scalable quantum network[15,20]. A high mechanical quality factor $Q$ is an important figure of merit for all of these applications and single-crystal diamond is expected to have low dissipation due to its tightly bound carbon lattice, good tribological properties, and a stable, inert, and biocompatible surface[21].

Despite their promise, there have been few demonstrations of single-crystal diamond MEMs/NEMs for two primary reasons: first, heteroepitaxial growth of SCD is extremely difficult[22] and second, processing of SCD is not well developed. These two factors conspire to make thin (~1 μm) films of SCD challenging to produce reliably. Hence diamond resonators have been largely limited to polycrystalline diamond (PCD)[23,24], which can be heteroepitaxially grown on silicon and $SiO_2$[25]. However, with its many grains and grain boundaries, PCD suffers from higher internal loss (limiting quality factors) and inferior optical properties compared to SCD. Also, diamond defect spin centers such as NV centers have longer quantum coherence times in SCD[26], making it favorable over PCD for quantum applications in which a mechanical mode of the resonator is coupled to a spin[15,18]. A promising subsurface ion implantation technique[27] has recently been used to generate thin membranes of SCD[28]. But the high-energy ions used to form the membrane damage the lattice, compromising the mechanical and optical properties of diamond.

We have developed a gentle, robust, and reliable process to prepare thin films of high quality SCD, which we use to fabricate SCD MEMs. Using an oxide bonding technique[29], we form a diamond/$SiO_2$/silicon multilayer structure: a diamond-on-insulator (DOI) analogy to silicon on insulator (SOI). A cross section of the layered structure is shown in Fig. 1(a). The diamond device layer can be tailored to <1 micron and importantly, undergoes minimal treatment. These DOI structures provide a robust starting point for a variety of thin film diamond applications, such as mechanical resonators and diamond photonics. In this paper, we demonstrate the fabrication and measurement of high $Q$ SCD cantilevers from the DOI. The cantilevers have reproducibly high mechanical quality factors, a testament to the fact that we induce minimal damage to the intrinsic diamond structure.

To make the DOI structure, we started with commercial electronic grade SCD plates (Applied Diamond), with dimensions 2 mm x 2mm x 0.02 mm. A 30 nm layer of $SiO_2$ was deposited on the diamond plate using plasma-enhanced chemical vapor deposition (PECVD) at 250° C. The diamond plate was then bonded to an oxidized silicon chip using a low temperature oxide bonding technique[29]. Key to the success of the bonding is a smooth (< 1 nm surface roughness) and clean surface. The bond is robust and survives several subsequent processing steps, including intense ultrasounding in organic solvents and acids. The diamond device layer was etched approximately 18 μm to the desired thickness of 1-2 μm using an $ArCl_2$ RIE/ICP (reactive ion etch/inductively coupled plasma) etch. The $ArCl_2$ etch leaves a very smooth diamond surface[30], ~ 0.3 nm root-mean-square surface roughness as measured with an atomic force microscope.

The SCD cantilevers are photolithographically patterned on the DOI structure and etched with an $O_2$ plasma. A series of patterned cantilevers of varying dimensions is shown in Fig. 1(b). The cantilevers are released in a buffered HF acid etch that selectively removes the 1 μm $SiO_2$ layer underneath the cantilevers. A scanning electron micrograph (SEM) image of a released cantilever is shown in Fig. 1(c).

We characterized the mechanical properties of eight cantilevers under high vacuum at room temperature. The results as well as cantilever dimensions are presented in Table 1. The cantilever widths and lengths are as lithographically defined and the thicknesses are calculated from the cantilever's resonant frequency, using the Young's modulus and density of bulk diamond. These inferred thicknesses are within 20% of those measured using SEM images as well as interference fringe patterns in optical images. The cantilever thickness varies uniformly across the chip due to a wedge shape of the starting diamond plate, and varies by ~ 100nm across each cantilever.

We measured the resonant frequency $f_0$ and quality factor $Q$ by mechanically driving the cantilever with piezoelectric actuation and monitoring the position response of the cantilever with optical interferometric detection. A 532 nm wavelength laser is focused to a ~ 4 μm spot size onto the cantilever and interference occurs between the light reflected from the diamond cantilever and the underlying silicon substrate. The laser power incident on the cantilever is ~ 1 μW. An amplified photodetector records the intensity of the reflected beam and its output is sent to a lockin amplifier.

Figure 2(a) shows the frequency response of cantilever M7's amplitude of motion as the drive is swept through resonance at T = 293 K. As expected for a harmonic oscillator, the data follow a square-root of Lorentzian profile, centered on $f_0$ = 728.217 kHz. We measured the $Q$ factor through the characterstic ringdown time of the cantilever $\tau$ and the relation $Q = \pi \tau f_0$. Figure 2(b) shows the decaying amplitude of cantilever motion after the cantilever drive is turned off. An exponential fit to the data yields $\tau$ = 145 ms and $Q$ = 338,000. This $Q$ value is in good agreement with the $Q$ = 319,000 extracted from the full width, half maximum of the frequency sweep. As the ringdown measurement is a more accurate measure of the real $Q$ [31], all cited $Q$ values in this letter will be extracted from an average of 50 ringdown measurements. All eight measured cantilevers have reproducibly high $Q$, ranging between 53,600 and 338,000, as shown in Table 1. We note that initial measurements of $Q$ before careful cleanings were ~ 50-75% lower than the values in the table, which were measured after two rounds of the following three steps: piranha clean (5:1 $H_2SO_4$:$H_2O_2$ at 110° C, 5 min), an RCA1 clean (1:1:5 $NH_4OH$:$H_2O_2$:$H_2O$ at 80° C, 5 min), and an RCA2 clean (1:1:5 $HCl$:$H_2O_2$:$H_2O$

at 80° C, 5 min). We measured the *Q*'s before and after each round of cleaning, and in almost all cases found *Q* to increase with each cleaning step. One exception was cantilever M1, whose *Q* dropped from 53,600 to 22,000 after the second clean. We also observed subsequent drops in *Q* for a few of the resonators after prolonged exposure to air. High-*Q* resonators are particularly susceptible to surface cleanliness and we believe that random surface contamination is responsible for the observed changes in *Q* with cleaning and air exposure. To report data that is most indicative of intrinsic diamond device performance, all values of *Q* and $f_0$ in Table 1 are measured directly after the second cleaning, with the exception of M1, which was measured after the first cleaning.

The strength of our measurement can affect apparent *Q*-values; in particular laser-induced photothermal effects can drive or damp mechanical oscillators[32] and strong mechanical drives can induce mechanical nonlinearities. We carefully studied the effects of laser power and drive amplitude on our measured *Q*'s and found that above certain threshold values, they could enhance or reduce the apparent *Q*. For all the data presented here, we selected laser powers and piezo drive voltages well below these threshold values to operate in a regime insensitive to their effect.

Dissipation mechanisms in mechanical resonators are often temperature dependent[33]. To shed light on the dissipation mechanisms in SCD, we measured the *Q* of three resonators as a function of temperature from room temperature down to 10 K in a variable temperature $^4$He cryostat (Montana Instruments). The results plotted in Fig. 3 show a strong nonmonotonic dependence on temperature, with all three cantilevers showing similar behavior. As temperature is lowered from 293 K to ~ 120-150K, *Q* slowly rises and reaches a maximum as high as 205,000. Then *Q* sharply drops as temperature is lowered to ~50K and begins to rise again for lower temperatures. The temperature data presented here were taken after only one round of cantilever cleaning and hence the room temperature *Q*'s were lower than those presented in Table 1. The highest *Q* measured was 205,000 at T = 160 K in cantilever M2, whose room temperature *Q* was 76,400. We note that the temperature dependence observed in our measurements is qualitatively very different than previously reported measurements on PCD resonators[24] and SCD resonators formed by subsurface ion implantation[34].

The excellent room temperature and low temperature *Q*-values reported here confirm predictions that SCD is a promising material for MEMs applications. The *Q*'s are an order of magnitude higher than those in SCD resonators produced by the subsurface ion implantation technique[34,35] and in PCD resonators[23,24]. For force sensing applications a cantilever's sensitivity is ultimately limited by its Brownian motion, with a thermally limited force sensitivity $F_{\min} = \sqrt{\frac{2kk_BTB}{\pi f_0 Q}}$ where *k* is the cantilever spring constant, $k_B$ is the Boltzmann constant, and *B* is the measurement bandwidth. Cantilever M4 has an $F_{\min} = 4\times10^{-16} N/Hz^{1/2}$ at room temperature and an $F_{\min} = 1\times10^{-16} N/Hz^{1/2}$ at 10K. We attribute the high quality factors to a combination of diamond's low intrinsic loss, inert surface, and the nondestructive process with which we form a thin device layer of diamond.

The dissipation mechanisms in our SCD resonators are not fully understood and we discuss a few possibilities here. Viscous damping due to collisions with air molecules is negligible at our operating pressures. We measured pressure dependent *Q*'s at room temperature and found them to saturate below ~ 0.1 mTorr. Thermoelastic dissipation sets an upper bound on *Q* of $Q_{TED}$ ~ $10^7$ for the range of cantilevers we have studied[36], approximately two orders of magnitude higher than our observations. Diamond's high thermal conductivity, which makes the thermal time constant across the diamond resonator much shorter than the cantilever oscillation period, is responsible for this large value of $Q_{TED}$. Clamping losses are geometry dependent, setting an upper limit of $Q_{clamping} = 2.17\frac{l^3}{t^3}$ where $l$ is the cantilever length, and $t$ is cantilever thickness[37]. Figure 4 shows the calculated $Q_{clamping}$ for all eight cantilevers as well as the measured $Q$'s before the first cleaning ($Q_1$), after the first cleaning ($Q_2$), and after the second cleaning ($Q_3$). The ratio of $Q_{clamping}$ to our observed $Q$ values, $r_c = \frac{Q_{clamping}}{Q}$, is in the range of 1-4 for the eight resonators, with $Q$ taken to be $Q_3$. At T = 293 K four of the cantilevers show $r_c \sim 1$ (M2: $r_c = 1.3$, M5: $r_c = 0.9$, M6: $r_c = 1.2$, M7: $r_c = 0.8$), indicating that clamping loss may be a relevant dissipation mechanism in some of our resonators. The $Q$ of cantilevers M5 and M7 even exceeds $Q_{clamping}$, which may be explained by the uncertainty in the estimated thickness. Though we observed no clear dependence on surface-volume ratio, which varies by a factor of two over the cantilevers, we did see a significant improvement in $Q$ over two rounds of cleaning and subsequent drops in $Q$ after prolonged air exposure. It is likely then that surface dissipation plays a role in some of the cantilevers. The nonmonotonic temperature dependence that has a broad minimum in $Q$ at ~ 50K could be attributed to an internal friction peak, such as those that have been observed in silicon at temperatures ~ 135 K[33], or to surface imperfections.

Figure 4 summarizes the discussion of dissipation in our SCD resonators: we have evidence that a combination of clamping losses and surface losses are limiting the $Q$'s near room temperature. These results are encouraging as both of these loss mechanisms can be mitigated through cantilever design and processing. For singly clamped beams, reducing $\frac{l^3}{t^3}$ will minimize clamping loss. More exotic cantilever geometries that minimize clamping loss have recently been explored theoretically and experimentally[38] and these geometries can be fabricated in SCD starting from our DOI platform. Together with careful surface preparation techniques, we are optimistic that higher *Q*'s can be achieved in SCD resonators.

We have demonstrated that single crystal diamond mechanical resonators exhibit excellent mechanical quality factors, leading to high force sensitivity and exciting prospects for integration of SCD into MEMs based devices. In the future, we envision coupling an intrinsic diamond spin, such as an NV center, to the mechanical mode of the oscillator as a means of incorporating NV centers into a quantum network. Additionally we have developed a nondestructive process to form thin films of high quality diamond in a DOI structure, which provides a general-purpose starting platform for diamond-based MEMs, photonics, and microelectronics. Importantly, our results show that our SCD MEMs are not yet at their intrinsic dissipation limit, and we see a viable path to even higher performance SCD resonators diamond based MEMs/NEMs using our DOI structure.

**Acknowledgements:** This work was funded by the AFOSR and DARPA. A portion of the work was done in the UC Santa Barbara nanofabrication facility, part of the NSF funded NNIN network.

## References:

1. Coe, S. & Sussmann, R. Optical, thermal and mechanical properties of CVD diamond. *Diamond and Related Materials* **9**, 1726–1729 (2000).
2. Walker, J. Optical absorption and luminescence in diamond. *Reports on Progress in Physics* **42**, 1605 (1979).
3. *The Properties of Diamond*. (Academic Press: London, 1979).
4. Isberg, J. High Carrier Mobility in Single-Crystal Plasma-Deposited Diamond. *Science* **297**, 1670–1672 (2002).
5. Lee, K. C. *et al.* Entangling Macroscopic Diamonds at Room Temperature. *Science* **334**, 1253–1256 (2011).
6. Neumann, P. *et al.* Multipartite Entanglement Among Single Spins in Diamond. *Science* **320**, 1326–1329 (2008).
7. Hanson, R. & Awschalom, D. D. Coherent manipulation of single spins in semiconductors. *Nature* **453**, 1043–1049 (2008).
8. Childress, L., Gurudev Dutt, M., Taylor, J. & Zibrov, A. Coherent dynamics of coupled electron and nuclear spin qubits in diamond. *Science* **314**, 281 (2006).
9. Epstein, R. J., Mendoza, F. M., Kato, Y. K. & Awschalom, D. D. Anisotropic interactions of a single spin and dark-spin spectroscopy in diamond. *Nat Phys* **1**, 94–98 (2005).
10. Childress, L. *et al.* Coherent Dynamics of Coupled Electron and Nuclear Spin Qubits in Diamond. *Science* **314**, 281–285 (2006).
11. Rugar, D., Budakian, R., Mamin, H. & Chui, B. Single spin detection by magnetic resonance force microscopy. *Nature* **430**, 329–332 (2004).
12. Ekinci, K. L., Huang, X. M. H. & Roukes, M. L. Ultrasensitive nanoelectromechanical mass detection. *Appl. Phys. Lett.* **84**, 4469 (2004).
13. Thompson, J. D. *et al.* Strong dispersive coupling of a high-finesse cavity to a micromechanical membrane. *Nature* **452**, 72–75 (2008).
14. Chan, J. *et al.* Laser cooling of a nanomechanical oscillator into its quantum ground state. *Nature* **478**, 89–92 (2011).
15. Kolkowitz, S. *et al.* Coherent Sensing of a Mechanical Resonator with a Single-Spin Qubit. *Science* **335**, 1603–1606 (2012).
16. O'Connell, A. *et al.* Quantum ground state and single-phonon control of a mechanical resonator. *Nature* **464**, 697–703 (2010).
17. Pályi, A., Struck, P., Rudner, M., Flensberg, K. & Burkard, G. Spin-Orbit-Induced Strong Coupling of a Single Spin to a Nanomechanical Resonator. *Phys. Rev. Lett.* **108**, (2012).
18. Wilson-Rae, I., Zoller, P. & Imamoḡlu, A. Laser cooling of a nanomechanical resonator mode to its quantum ground state. *Phys. Rev. Lett.* **92**, 75507 (2004).
19. Maze, J. *et al.* Nanoscale magnetic sensing with an individual electronic spin in diamond. *Nature* **455**, 644–647 (2008).
20. Rabl, P. *et al.* A quantum spin transducer based on nanoelectromechanical resonator arrays. *Nat Phys* **6**, 602–608 (2010).
21. Popov, C. *et al.* Bioproperties of nanocrystalline diamond/amorphous carbon composite films. *Diamond and Related Materials* **16**, 735–739 (2007).
22. Schreck, M., Hörmann, F., Roll, H., Lindner, J. K. N. & Strizker, B. Diamond nucleation on iridium buffer layers and subsequent textured growth: A route for the realization of single-crystal diamond films. *Appl. Phys. Lett.* **78**, 192 (2001).
23. Adiga, V. *et al.* Mechanical stiffness and dissipation in ultrananocrystalline diamond microresonators. *Phys. Rev. B* **79**, (2009).
24. Hutchinson, A. B. *et al.* Dissipation in nanocrystalline-diamond nanomechanical resonators. *Appl. Phys. Lett.* **84**, 972 (2004).
25. Gsell, S., Bauer, T., Goldfuß, J., Schreck, M. & Strizker, B. A route to diamond wafers by epitaxial

deposition on silicon via iridium/yttria-stabilized zirconia buffer layers. *Appl. Phys. Lett.* **84**, 4541 (2004).
26. Casabianca, L. B., Shames, A. I., Panich, A. M., Shenderova, O. & Frydman, L. Factors Affecting DNP NMR in Polycrystalline Diamond Samples. *J. Phys. Chem. C* **115**, 19041–19048 (2011).
27. Parikh, N. R. *et al.* Single-crystal diamond plate liftoff achieved by ion implantation and subsequent annealing. *Appl. Phys. Lett.* **61**, 3124 (1992).
28. Aharonovich, I. *et al.* Homoepitaxial Growth of Single Crystal Diamond Membranes for Quantum Information Processing. **24**, 54 (2012).
29. Liang, D. *et al.* Low-Temperature, Strong SiO2-SiO2 Covalent Wafer Bonding for III–V Compound Semiconductors-to-Silicon Photonic Integrated Circuits. *Journal of Elec Materi* **37**, 1552–1559 (2008).
30. Lee, C. L., Gu, E., Dawson, M. D., Friel, I. & Scarsbrook, G. A. Etching and micro-optics fabrication in diamond using chlorine-based inductively-coupled plasma. *Diamond and Related Materials* **17**, 1292–1296 (2008).
31. Stipe, B., Mamin, H., Stowe, T., Kenny, T. & Rugar, D. Noncontact Friction and Force Fluctuations between Closely Spaced Bodies. *Phys. Rev. Lett.* **87**, (2001).
32. Metzger, C. H. & Karrai, K. Cavity cooling of a microlever. *Nature* **432**, 1002–1005 (2004).
33. Yasumura, K. Y. *et al.* Quality factors in micron-and submicron-thick cantilevers. *Microelectromechanical Systems, Journal of* **9**, 117–125 (2000).
34. Zalalutdinov, M. K. *et al.* Ultrathin Single Crystal Diamond Nanomechanical Dome Resonators. *Nano Lett.* **11**, 4304–4308 (2011).
35. Liao, M., Li, C., Hishita, S. & Koide, Y. Batch production of single-crystal diamond bridges and cantilevers for microelectromechanical systems. *J. Micromech. Microeng.* **20**, 085002 (2010).
36. Zener, C. Internal Friction in Solids. 1. Theory of Internal Friction in Reeds. *Phys.Rev.* **52**, 230–235 (1937).
37. Hosaka, H., Itao, K. & Kuroda, S. Damping characteristics of beam-shaped micro-oscillators. *Sensors and Actuators A: Physical* **49**, 87–95 (1995).
38. Cole, G. D., Wilson-Rae, I., Werbach, K., Vanner, M. R. & Aspelmeyer, M. Phonon-tunnelling dissipation in mechanical resonators. *Nature Communications* **2**, 231 (2011).

**Figures:**

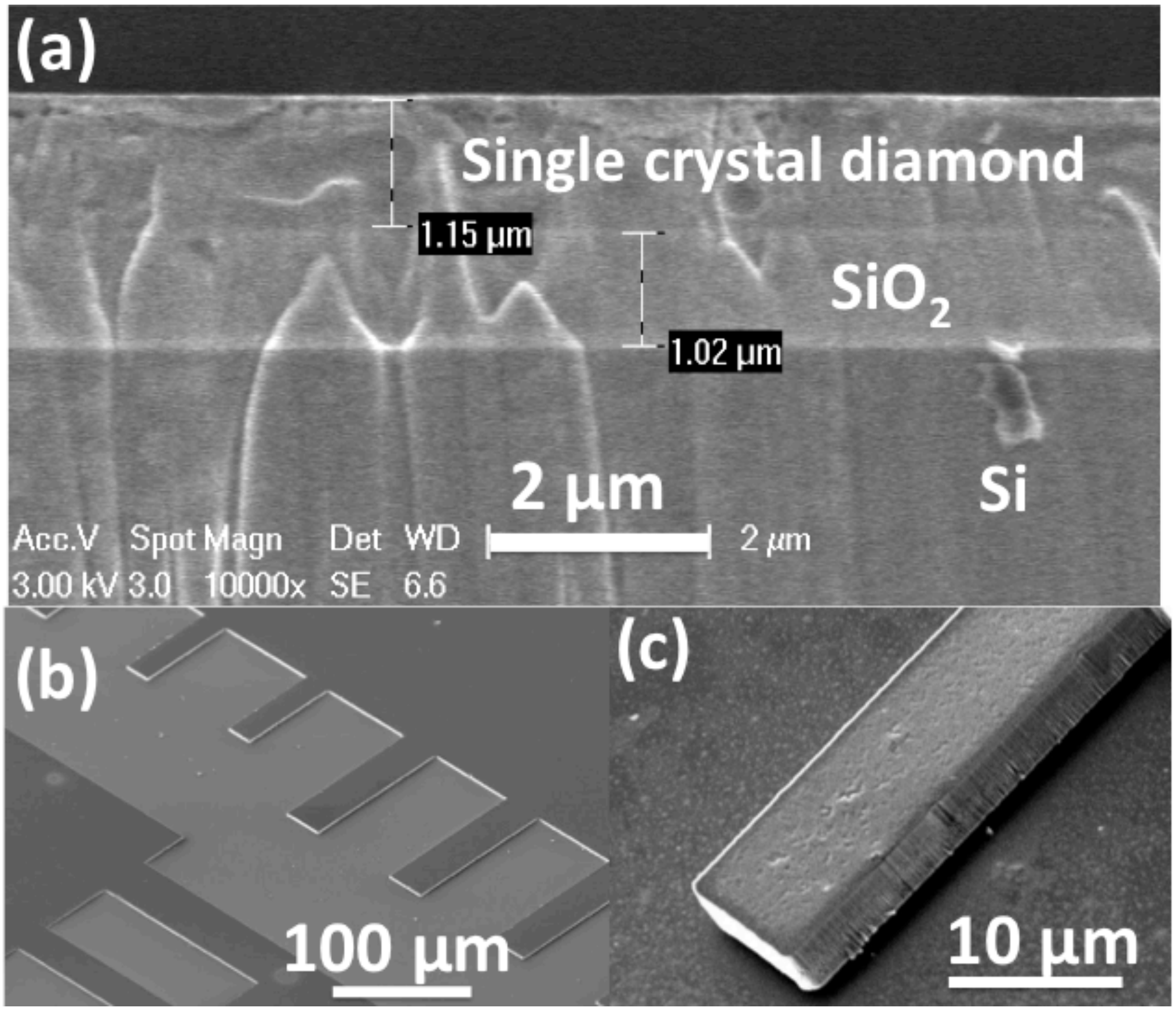

**Figure 1**: (a) SEM image of diamond-on-insulator cross section. Horizontal layers from top to bottom: 1.15 μm single-crystal diamond/1.02 μm $SiO_2$/ 500 μm Si. (b) SEM of patterned and etched single-crystal diamond cantilevers lying atop $SiO_2$. (c) Released single-crystal diamond cantilever suspended over Si.

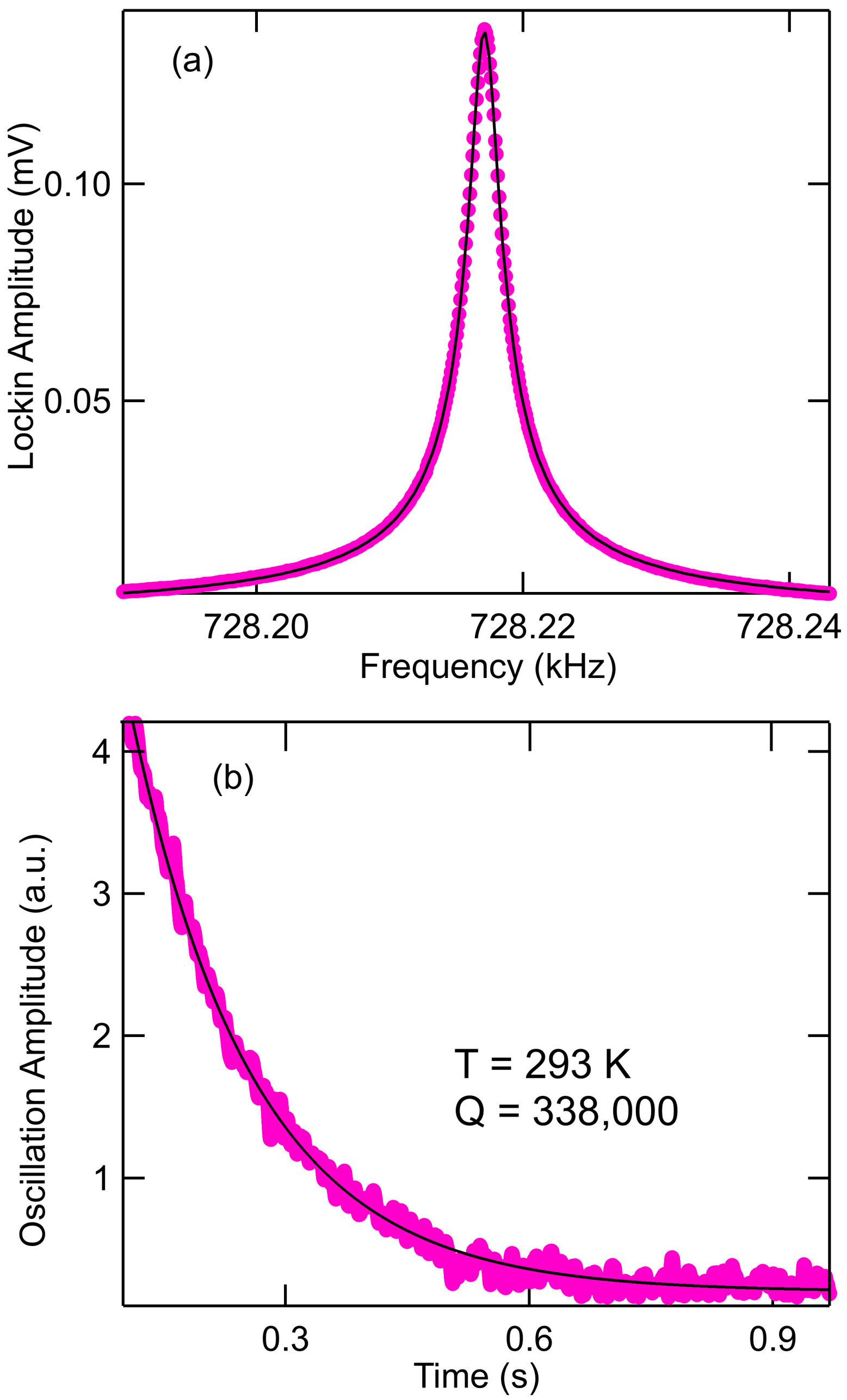

**Figure 2**: Cantilever M7 mechanical properties: (a) Cantilever motional amplitude as a function of drive frequency. (b) Cantilever mechanical ringdown at room temperature.

| Cantilever | length (µm) | width (µm) | thickness (µm) | $f_0$ (kHz) | *Q* |
|---|---|---|---|---|---|
| M1 | 60 | 15 | 1.7 | 1394.6 | 54,000 |
| M2 | 60 | 15 | 1.4 | 1146.5 | 129,000 |
| M3 | 60 | 15 | 1.1 | 890.81 | 128,000 |
| M4 | 60 | 15 | 0.76 | 637.58 | 222,000 |
| M5 | 80 | 20 | 2.1 | 1011.9 | 133,000 |
| M6 | 80 | 20 | 1.8 | 843.99 | 165,000 |
| M7 | 80 | 20 | 1.5 | 728.23 | 338,000 |
| M8 | 80 | 20 | 1.2 | 586.09 | 100,000 |

**Table 1**: Room temperature quality factor, resonant frequency, and dimensions of eight fabricated single-crystal diamond cantilevers.

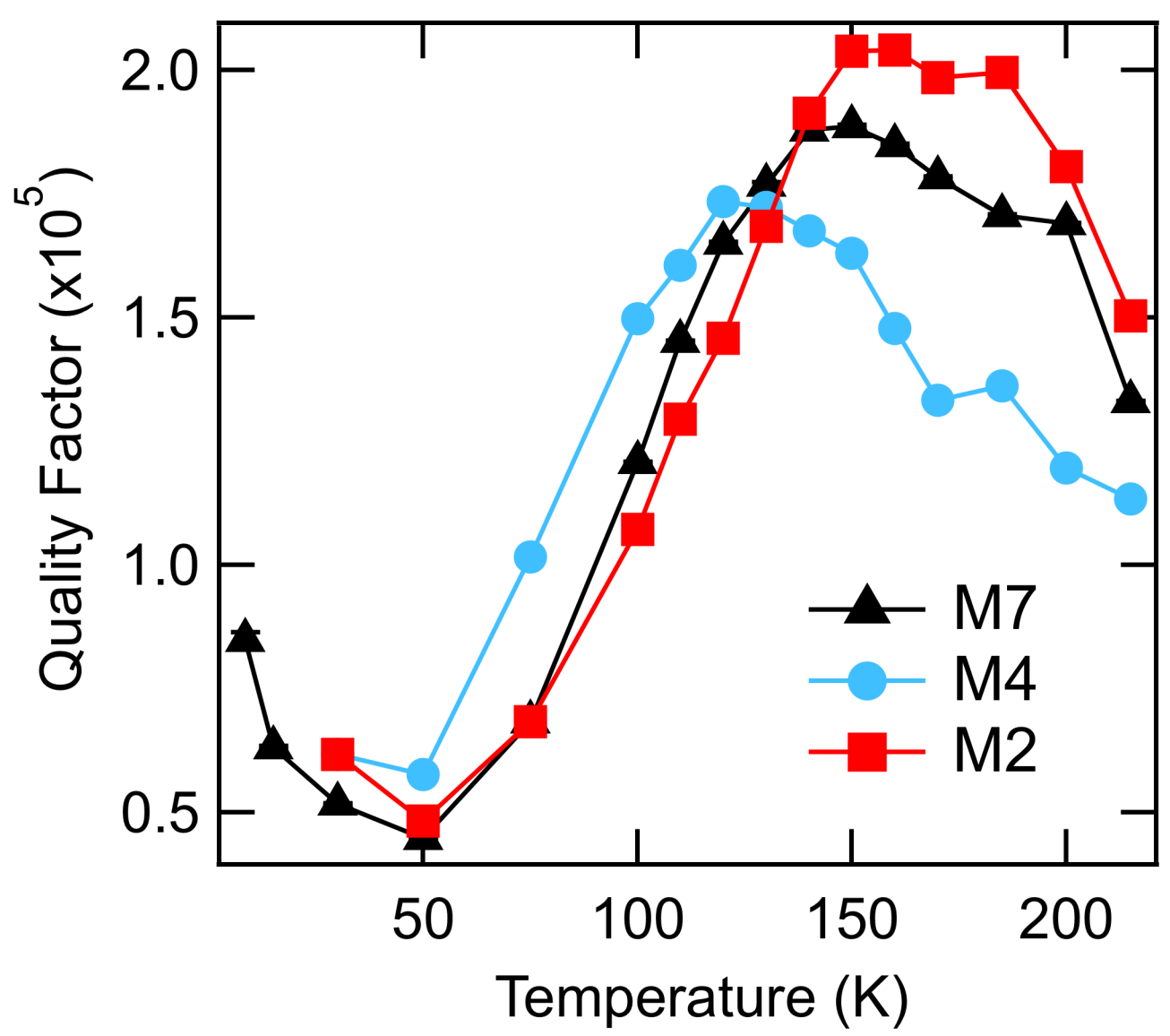

**Figure 3**: Temperature dependence of mechanical quality factors for three cantilevers: M2 (red squares), M4 (blue circles), M7 (black triangles). Error bars are smaller than the marker size.

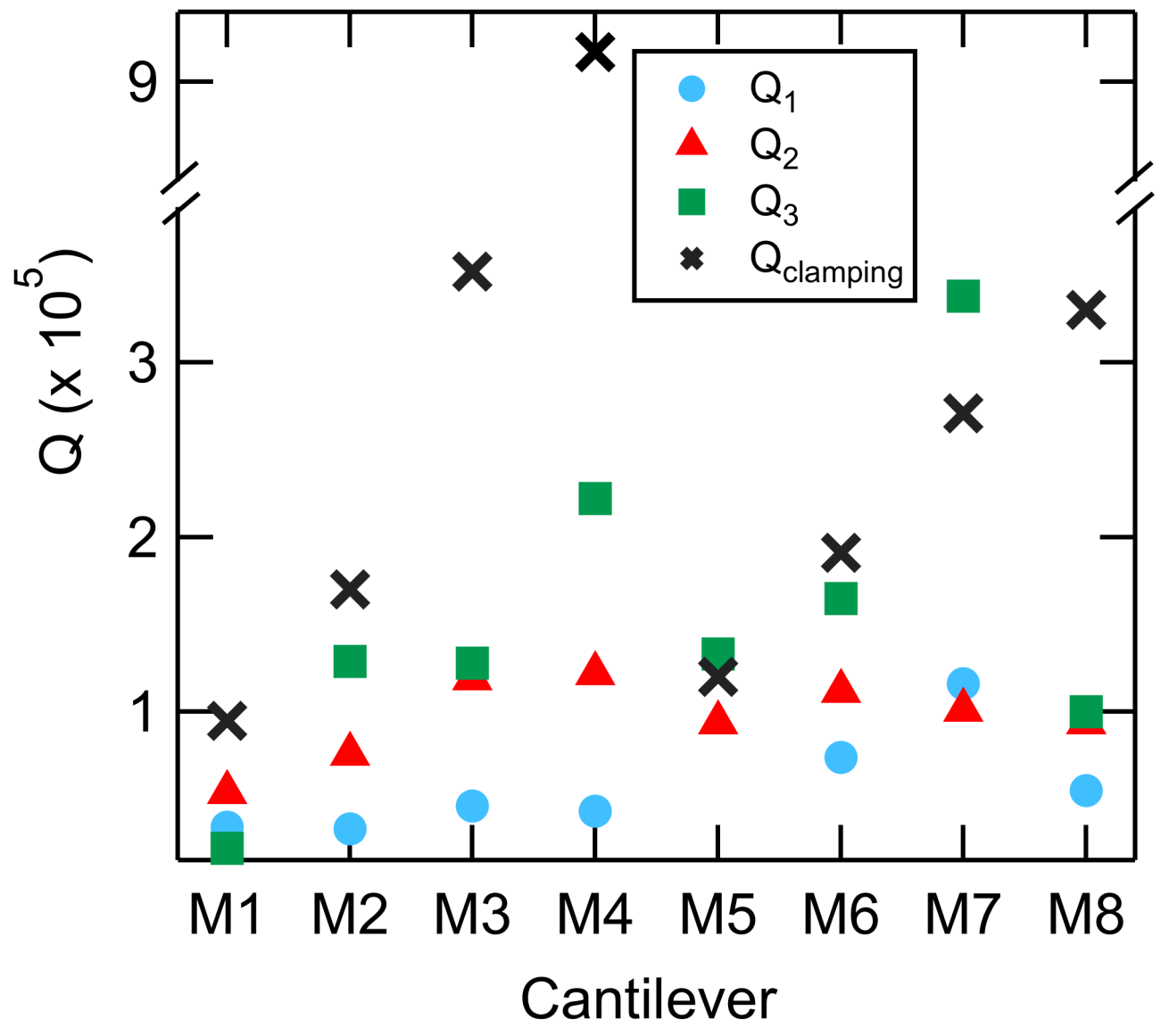

**Figure 4**: Quality factors for cantilevers M1–M8: before first cleaning ($Q_1$, blue circles), after first cleaning ($Q_2$, red triangles), and after second cleaning ($Q_3$, green squares). Plotted also are clamping loss limited $Q$'s (black x's), calculated using the dimensions of each cantilever. Note the break in the y-axis.